\documentclass[12pt, letterpaper]{article}

\usepackage[noline, boxed]{algorithm2e}
\usepackage{amsmath}
\usepackage{amssymb}
\usepackage{amsthm}
\usepackage{bbm}
\usepackage{bm}
\usepackage{booktabs}
\usepackage[font = footnotesize]{caption}
\usepackage{dsfont}
\usepackage{enumitem}
\usepackage[margin = 1in, bottom = 1.25in]{geometry}
\usepackage{graphicx}
\usepackage[colorlinks = true, linkcolor = blue, citecolor = red]{hyperref}
\usepackage{subcaption}
\usepackage{upgreek}
\usepackage{xcolor}

\newcommand{\fp}[1]{\left(#1\right)}

\newcommand{\fb}[1]{\left\{#1\right\}}

\newcommand{\indf}{\mathbbm{1}} 

\newcommand{\bth}{\bm{\theta}}

\newcommand{\bO}{\mathcal{O}}
\newcommand{\bS}{\bm{\Sigma}}

\usepackage{newtxtext} 

\usepackage[style = apa, backend = biber]{biblatex}
\DeclareLanguageMapping{american}{american-apa}
\title{High-Resolution Retrieval of Atmospheric Boundary Layers with
Nonstationary Gaussian Processes}
\author{
  Haoran Xiong\thanks{Corresponding author: \texttt{haoran.xiong@wisc.edu}}
  \thanks{Department of Statistics, University of Wisconsin-Madison}
  \and
  Paytsar Muradyan\thanks{Environmental Science Division, Argonne National Laboratory}
  \and
  Christopher J. Geoga\footnotemark[2]
}
\date{}

\begin{document}

\maketitle

\begin{abstract}
    The atmospheric boundary layer (ABL) plays a critical role in governing 
turbulent exchanges of momentum, heat moisture, and trace gases between 
the Earth's surface and the free atmosphere, thereby influencing 
meteorological phenomena, air quality, and climate processes. 
Accurate and temporally continuous characterization of the ABL structure 
and height evolution is crucial for both scientific understanding and 
practical applications.
High-resolution retrievals of the ABL height from vertical velocity 
measurements is challenging because it is often estimated using empirical 
thresholds applied to profiles of vertical velocity variance or related 
turbulence diagnostics at each measurement altitude, which can suffer from 
limited sampling and sensitivity to noise. 
To address these limitations, this work employs nonstationary Gaussian
process (GP) modeling to more effectively capture the spatio-temporal 
dependence structure in the data, enabling high-quality---and, if desired,
high-resolution---estimates of the ABL height without reliance on ad-hoc
parameter tuning. 
By leveraging Vecchia approximations, the proposed method can be applied to 
large-scale datasets, and example applications using full-day vertical 
velocity profiles comprising approximately $5$M measurements are 
presented.

\end{abstract}


\newpage

\section{Introduction}\label{sec:intro}
The atmospheric boundary layer (ABL), also known as the planetary boundary 
layer (PBL), forms the lowest part of the atmosphere and is directly 
influenced by surface processes such as momentum flux, sensible and latent 
heat exchange, and surface radiative and moisture variations~\parencite{stull2012, garratt1994}. 
Its depth typically ranges from several hundred meters to a few kilometers, 
depending on the time of day, surface characteristics, and synoptic 
conditions. 
During daytime, solar radiation heats the surface, enhancing buoyant 
convection and turbulence, which deepen the ABL. 
At night, radiative cooling and reduced turbulent mixing produce a stably 
stratified and shallow ABL~\parencite{stull2012, mahrt2014stably}. 
Capturing these diurnal and synoptic-scale variations is crucial for 
weather forecasting, air-quality assessment~\parencite{holtslag1993local, dabberdt2004meteorological}, 
and climate studies~\parencite{holtslag2013stable, baklanov2014online}.

A common challenge in ABL research is to estimate the ABL height 
using remote-sensing observations such as Doppler Lidar (DL)~\parencite{geoga2021, wang2021comprehensive, pearson2009DL, tucker2009},
micropulse LIDAR~\parencite{dang2019MPL, wang2021MPL}, or ceilometer backscatter 
profiles \parencite{eresmaa2006CL,kotthaus2020CL}. 
For DLs, the ABL height is often inferred from vertical velocity variance or 
spectral width profiles, which serve as proxies for turbulence intensity. 
Current methods involve dividing the data into fixed time segments at each 
altitude, computing marginal variances while assuming temporal independence, 
and applying empirical thresholds to identify the transition from turbulent 
(mixed-layer) to laminar (free-atmosphere) regimes~\parencite{tucker2009}. 
These traditional variance-threshold methods, however, have inherent 
limitations arising from the intermittent nature of atmospheric turbulence. 
Because turbulent motions are highly variable in space and time, segments 
of low-variance can occur within the ABL, while elevated variance values 
may also appear above it, particularly in the presence of cloud layers, 
shear zones, or residual-layer turbulence. 
This intermittency and overlap between regimes can lead to substantial 
retrieval uncertainty and frequent misidentification of the true ABL top. 
Ignoring temporal and spatial dependence also introduces bias-variance 
tradeoffs, which becomes particularly problematic under rapidly evolving 
ABL conditions.  
Moreover, a single universal variance threshold rarely performs consistently 
across different atmospheric stability regimes, cloud cover, or time of day, 
limiting automation of long-term retrieval without manual tuning. 

\begin{figure}[!ht]
    \centering
    \begin{subfigure}{0.8\textwidth}
        \centering
        \includegraphics[width=\linewidth]{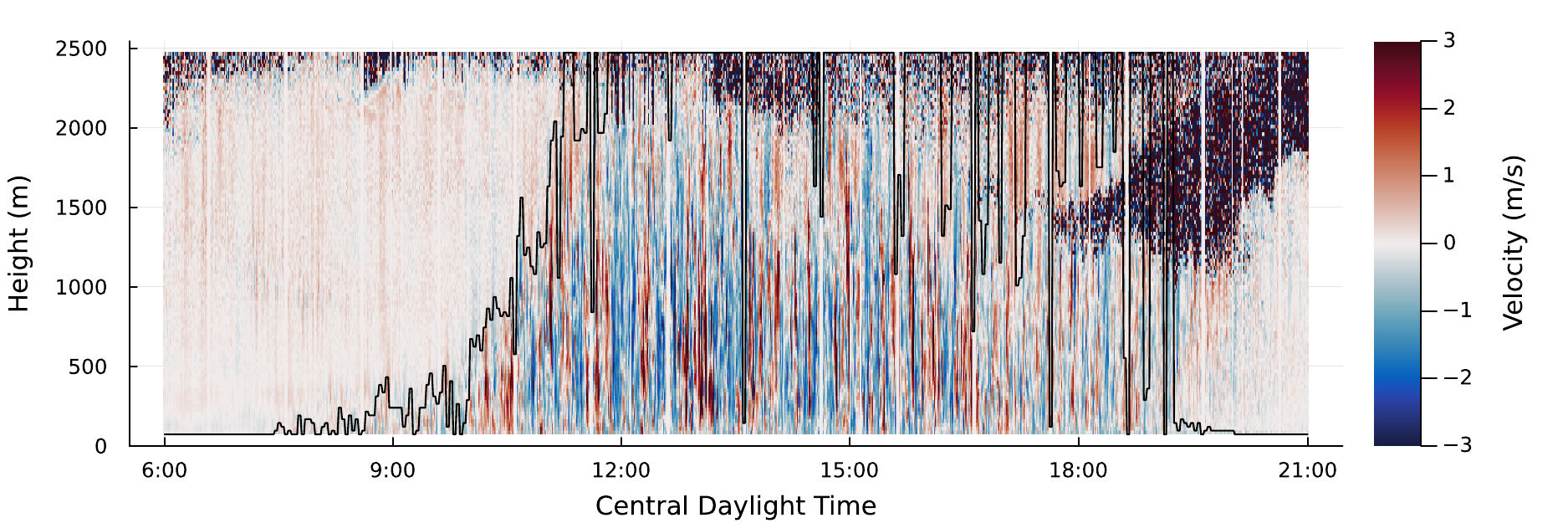}
        \caption{The ABL height estimation for June 23, 2023.}
    \end{subfigure}

    \begin{subfigure}{0.8\textwidth}
        \centering
        \includegraphics[width=\linewidth]{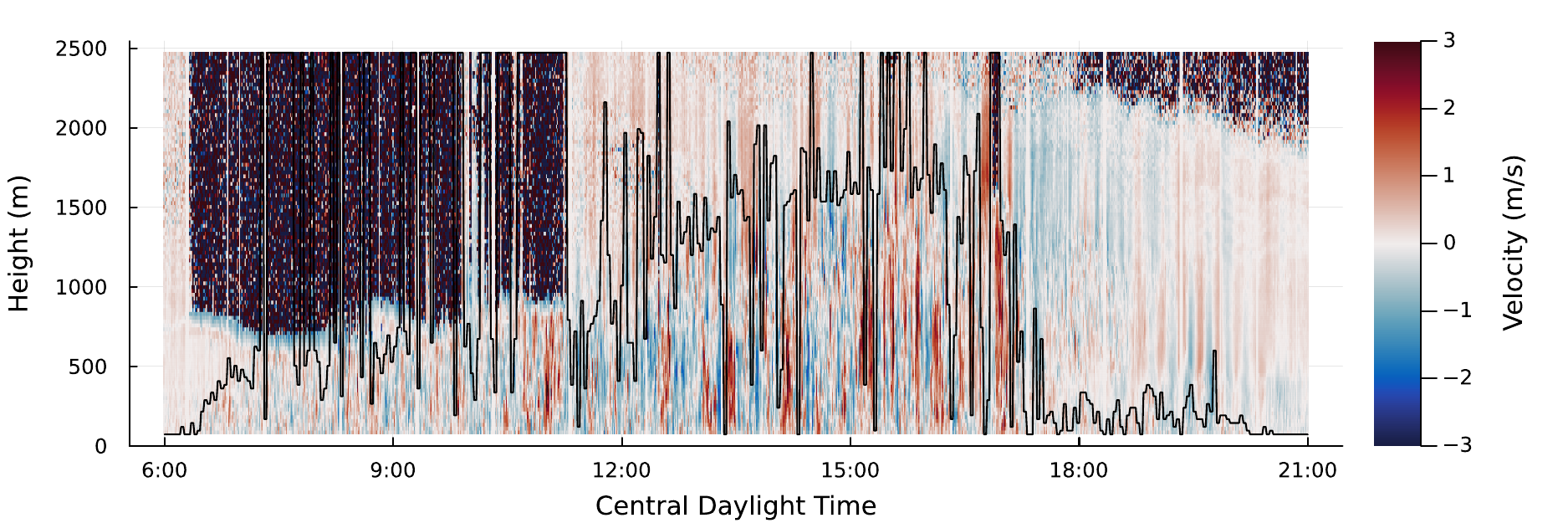}
        \caption{The ABL height estimation for June 27, 2023.}
    \end{subfigure}

    \begin{subfigure}{0.8\textwidth}
        \centering
        \includegraphics[width=\linewidth]{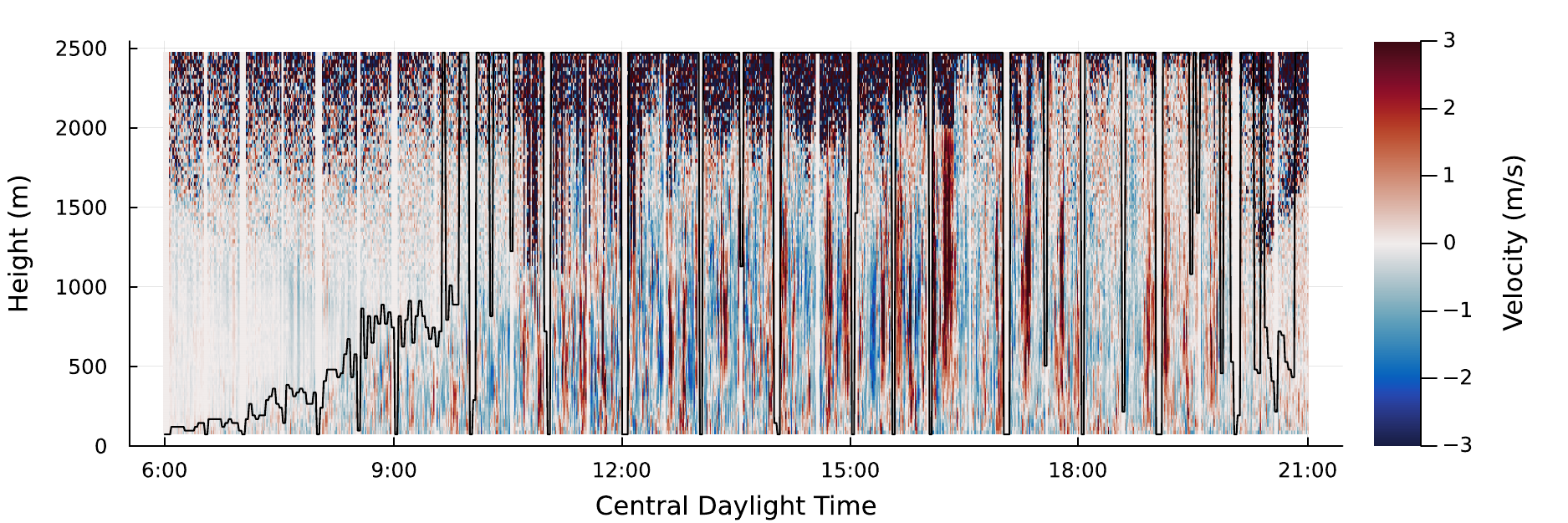}
        \caption{The ABL height estimation for July 20, 2023.}
    \end{subfigure}

    \caption{
        The ABL height estimation for three days with variance threshold 
        0.04 $m^2 \cdot s^{-2}$.
    }
    \label{fig:var_thres}
\end{figure}

To illustrate these challenges, Figure~\ref{fig:var_thres} shows an example 
ABL height estimation based on a variance threshold for three different 
days under different cloud conditions. 
In this example, the variance is calculated over windows of 300 samples 
that last approximately 160 seconds each, and a variance threshold of 0.04 
$m^2 \cdot s^{-2}$, following ~\textcite{tucker2009}, was applied.
The dark regions at higher altitudes correspond to low signal-to-noise 
ratio (SNR) and cloud contamination or blockage, where DL returns are 
unreliable.
As is clear from visual inspection, this threshold-based estimate 
fluctuates considerably. 
While temporal smoothing could reduce this variability, it would further 
degrade temporal resolution of the estimator, reducing the ability to 
resolve rapid features in the boundary-layer transition.

In this work, we propose an alternative approach using nonstationary 
Gaussian processes (GPs) models specifically designed to capture the strong 
spatio-temporal correlations inherent in vertical velocity measurements, 
which can be used to identify local changes of the ABL height. 
The key idea of our formulation is that a time-varying ABL height is embedded 
within the covariance function of the GP, so that the maximum likelihood 
estimation of the covariance parameters simultaneously yields the ABL height. 
By explicitly modeling cross-time and cross-height dependencies, the GP 
model extracts more information than a marginal variance-based approach 
and significantly mitigates the disadvantage of missing the correlation 
between contiguous time segments. 
Moreover, it enables continuous, high temporal resolution retrievals, 
approaching the native measurement frequency, and provides data-driven 
differentiation between turbulent and non-turbulent layers, eliminating 
the need for ad-hoc variance thresholds. 

The remainder of this paper is structured as follows. 
Section~\ref{sec:method} introduces the proposed methodology, detailing 
model assumptions, formulations, and computational framework. 
Section~\ref{sec:results}, presents results from three representative case 
studies under distinct cloud conditions. 
Finally, Section~\ref{sec:dis} discusses implications of our findings and 
outlines potential directions for future research.

\section{Methodology}\label{sec:method}

\subsection{Gaussian processes}\label{sec:GPs}

A Gaussian process (GP) is a stochastic process where every finite set of random
variables follows a multivariate Gaussian distribution. The process is typically
indexed by time or space and thus suitable for modeling in many environmental
applications, including wind profiles. Since a multivariate Gaussian distribution
only depends on its mean vector and covariance matrix, a GP is uniquely
determined by its mean and covariance function, making it one of the easiest
multivariate laws to fully specify. Practitioners often assume a zero mean,
use a parametric covariance function,  and estimate its parameters from a
collection of observations. Specifically, the distribution of $n$ observations
$\bm{y} = \{y_j\}_{j=1}^n$ at different locations $\{\bm{x}_j\}_{j=1}^n$ is
$\mathcal{N}(\bm{0}, \bm{\Sigma})$, where $\bm{\Sigma}$ is the covariance matrix
induced by the covariance function $K(\cdot,\cdot|\bm{\theta})$ with a parameter
vector $\bm{\theta}$ such that $\bm{\Sigma}_{j,k} = K(\bm{x}_j, \bm{x}_k|
\bm{\theta})$.  A common choice for estimating $\bm{\theta}$ is the maximum
likelihood estimator (MLE), which is the vector $\widehat{\bm{\theta}}$ that
maximizes the likelihood. For the convenience of optimization, people usually
consider minimizing the negative log-likelihood 
\begin{equation}\label{eq:nll} 
  -\ell(\bm{\theta}) = \frac{n}{2}\log(2\pi) + 
    \frac{1}{2} \log|\bm{\Sigma}(\bm{\theta})| + 
    \frac{1}{2} \bm{y}^T \bm{\Sigma}(\bm{\theta})^{-1} \bm{y}, 
\end{equation} 
where $|\bm{\Sigma}(\bm{\theta})|$ denotes the determinant of
$\bm{\Sigma}(\bm{\theta})$. One of the several significant conveniences of GP
models is that conditional expectations are linear transformations of the data
and conditional distributions are themselves Gaussian, making prediction and
uncertainty quantification simple and interpretable. They are also extremely
flexible interpolators because the choice of covariance function has a
significant impact on the behavior and implied conditional distributions of
predictions \parencite{stein1999}. By choosing one covariance function over
another, one can effectively determine the correlation length scale, (local)
anisotropy, mean-square differentiability (including fractional levels), and
more.

In this work in particular, the GP model will use the fact that local behavior
above and in the ABL differs in several important ways. Inspection of Figure
\ref{fig:var_thres} shows several immediate local properties that change sharply
with altitude: the spatial dependence length scale is significantly higher above
the ABL, and the temporal roughness and marginal variance is materially higher
in the ABL versus above it. More subtle instrument artifacts can also be
observed, such as higher measurement noise at higher altitudes. Encoding these
physical and measurement characteristics into the covariance function of a GP
model will mean that they can be utilized to accurately distinguish measurements
that are in the ABL versus above it.

As is common for contemporary datasets and problems, there are two challenges to
this GP application that require discussion: model design and computation. It is
clear that using an off-the-shelf covariance function will not be possible for
this application, and building a covariance function that has been specifically
designed to be fit-for-purpose will be necessary. The second issue is even more
universal: the computation of~\eqref{eq:nll} involves calculating a determinant
and solving a linear system, which for a dense $\bm{\Sigma}(\bth)$ is most
effectively done with a Cholesky factorization that requires $\mathcal{O}(n^3)$
work and $\mathcal{O}(n^2)$ storage. For reference, this work will be applied to
full-day profiles that contain approximately $5$M measurements. A square matrix
of edge length $5$M would require approximately $182$ {terabytes} of RAM to
store, and would obviously require an unacceptable amount of time to factorize
even once---let alone hundreds of times as would naturally be required in the
process of using an optimizer to compute an MLE. We will discuss design choices
to address both of these issues in the next two sections.

\subsection{A GP model for DL vertical velocity profiles}\label{sec:lidar_gp}

In the particular case of DL vertical velocity profiles, we adopt a
nonstationary GP model formulated as a mixture of two dependent (locally)
stationary components representing the turbulent (in-ABL) and laminar
(above-ABL) regimes.  This formulation extends the mixture-of-stationary
nonstationary modeling framework of~\textcite{fuentes2001b} to allow for
explicit dependence between the two regimes and smooth temporal variations of
their relative influence.  Specifically, within time intervals over which the
in-ABL and above-ABL dynamics can each be considered approximately stationary,
we consider the model 
\begin{equation}\label{eq:model}
    Z(t, x) = (1 - w(t, x)) Z_1(t, x) + w(t, x) Z_2(t, x),
\end{equation}
which can be treated as a convex combination of the two components of a
bivariate stationary Gaussian process with components denoted $Z_1$ and $Z_2$
that represent the processes in and above the ABL respectively. The function 
$w(t,x) \in (0, 1)$ here encodes information about the ABL height, up-
or down-weighting one term or the other in the process depending on whether a
given altitude $x$ is above the ABL or not. In particular, we use the simple
parameterization of
\begin{equation}\label{eq:weight}
    w(t, x) = \frac{1}{1 + \exp\{-s(x - \alpha(t))\}},
\end{equation}
where $\alpha(t)$ is the ABL height at time $t$, and $s$ is the parameter
adjusting the transition rate between $Z_1$ and $Z_2$. When the point
$(t,x)$ lies well below the ABL height, $w(t,x)$ is close to $0$, and $Z_1$ is the
dominant term in $Z$.  When the point $(t,x)$ lies well above the ABL, $w(t,x)$
is close to $1$, and $Z_2$ is dominant in $Z$. The ABL height $\alpha(t)$ is 
parameterized as
\begin{equation}\label{eq:height}
  \alpha(t) = \sum_{j=1}^m c_j \psi_j(t),
\end{equation}
where $\psi_j(t)$'s are local basis functions that can be placed as finely or
coarsely as desired for the specific application. In the next section, we will
demonstrate two different designs for $\alpha(t)$; one, based on wavelets, is
ideal for high-resolution estimation of the ABL height. The other, based on
splines, is useful at coarser timescales for producing full-day profiles of the
ABL, including in the presence of gaps and other irregularities in the data.

\begin{figure}[!ht]
  \centering
  \includegraphics[width=0.8\textwidth]{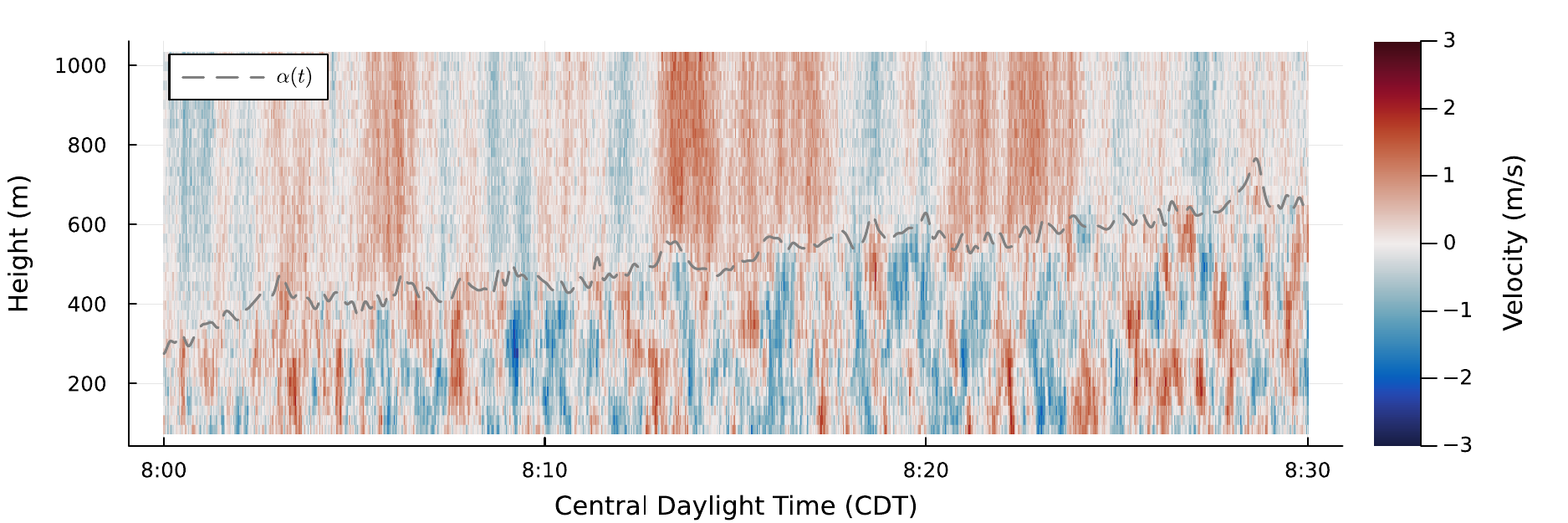}
  \caption{An example of simulated data from the ABL-mixture process given
  by~\eqref{eq:model}.}
  \label{fig:sim_demo}
\end{figure}

Using the multivariate Mat\'ern model of~\parencite{gneiting2010}, the marginal
covariance functions for $Z_1$ and $Z_2$ are given by 
\begin{equation} \label{eq:kernel_marginal}
  K_j\big((t,x), (t',x')| \bth_j \big) = \sigma_j^2 \frac{2^{1 - \nu_j}}{\Gamma(\nu_j)} (r_j \sqrt{2\nu_j})^{\nu_j} \mathcal{K}_{\nu_j}(r_j \sqrt{2\nu_j}) + \tau_j^2 \indf_{\{t=t', x=x'\}},
\end{equation}
for $j \in \{1,2\}$, where $\mathcal{K}_{\nu_j}(\cdot)$ denotes the modified
Bessel function of the second kind with order $\nu_j$~\parencite{nist}, $r_j$ is the
distance between $(t,x)$ and $(t',x')$ defined by 
\begin{equation} \label{eq:distance}
  r_j^2 = \frac{(t-t')^2}{(\rho_j^t)^2} + \frac{(x-x')^2}{(\rho_j^x)^2},
\end{equation}
and $\bth_j = (\sigma_j, \rho^t_j, \rho^x_j, \nu_j, \tau_j)$.  $\tau_j^2$ is the
variance of the nugget, i.e. we add to the variance at $(t,x)$ an extra variance
accounting for the measurement noise. The cross-covariance is given by
\begin{equation} \label{eq:kernel_cross}
  K_{12}\big((t,x), (t',x')| \bth_{12} \big) = 
  \beta \sigma_1 \sigma_2 \frac{2^{1 - \nu_{12}}}{\Gamma(\nu_{12})} 
  (r_{12} \sqrt{2\widetilde{\nu}_{12}}) \mathcal{K}_{\nu_{12}}(r_{12} \sqrt{2\widetilde{\nu}_{12}})
  \frac{\sqrt{\rho_1^t \rho_2^t}}{\rho_{12}^t}
  \frac{\sqrt{\rho_1^x \rho_2^x}}{\rho_{12}^x}
  \frac{\sqrt{\nu_1 \nu_2}}{\nu_{12}},
\end{equation}
where $\beta$ is the correlation coefficient, 
\begin{equation} \label{eq:param_cross}
  \nu_{12} = \frac{\nu_1 + \nu_2}{2},\ 
  \widetilde{\nu}_{12} = \frac{2 \nu_1 \nu_2}{\nu_1 + \nu_2},\ 
  (\rho_{12}^t)^2 = \frac{\nu_2 (\rho_1^t)^2 + \nu_1 (\rho_2^t)^2}{\nu_1 + \nu_2},\ 
  (\rho_{12}^x)^2 = \frac{\nu_2 (\rho_1^x)^2 + \nu_1 (\rho_2^x)^2}{\nu_1 + \nu_2}, 
\end{equation}
and 
\begin{equation} \label{eq:distance_cross}
  r_{12}^2 = \frac{(t - t')^2}{(\rho_{12}^t)^2} + \frac{(x - x')^2}{(\rho_{12}^x)^2}.
\end{equation}
The parameter vector of the cross-covariance is given by $\bth_{12} = (\beta,
\sigma_1, \rho_1^t, \rho_1^x, \nu_1, \sigma_2, \rho_2^t, \rho_2^x, \nu_2)$.  The
cross-covariance~\eqref{eq:kernel_cross} has a similar structure as the
marginal~\eqref{eq:kernel_marginal}, and the cross-covariance parameters are all
a certain kind of average of their marginal counterparts.
For example, $\widetilde{\nu}_{12}$ is the harmonic mean
of $\nu_1$ and $\nu_2$, and $\rho_{12}^t$ is a weighted quadratic mean of
$\rho_1^t$ and $\rho_2^t$.  It is worth noting that there are several different
multivariate Mat\'ern models in the literature, such as~\parencite{kleiber2012} and
~\parencite{yarger2024}. More generally, half-spectral frameworks
\parencite{stein2005b,horrell2017,geoga2021}
can be used to specify significantly broader and more flexible multivariate
laws. All of these alternatives could be substituted into this framework with
minimal modification, and we elect for the simplest choice here solely for
expediency. The most important aspect of this model and application is the
weight function $w(t, x)$, which encodes the ABL height. Figure~\ref{fig:sim_demo}
gives an example of a simulated ABL mixture model with parameters chosen
to emphasize the transition and dependence structure, demonstrating how
this particular design naturally creates sample paths that mimic the true data's
sharp transition in local behaviors.

Combining~\eqref{eq:model} with~\eqref{eq:kernel_marginal}
and~\eqref{eq:kernel_cross}, the marginal covariance function of $Z(t,x)$ is
given by 
\begin{equation} \label{eq:kernel_mixed}
  \begin{split}
    K\big( (t,x), (t',x') | \bth \big) = & \big(1 - w(t,x)\big) \big(1 - w(t',x')\big) K_1\big( (t,x), (t',x') | \bth_1 \big) \\
    & + w(t,x) w(t',x') K_2\big( (t,x), (t',x') | \bth_2 \big) \\
    & + \big(w(t,x) + w(t',x') - 2w(t,x)w(t',x')\big) K_{12}\big( (t,x), (t',x') | \bth_{12} \big),
  \end{split}
\end{equation}
where $\bth$ contains parameters from $K_1$, $K_2$, $K_{12}$, and $w(t,x)$.

\subsection{Fast likelihood approximation: Vecchia's method}\label{sec:vecchia}

Fast approximation of the Gaussian log-likelihood is an important and widely
studied problem in statistics and applied mathematics, and a variety of popular
methods have been proposed and analyzed. Broadly speaking, for covariance
function-based models these approaches fall into two categories: approximating
$\bS$ or approximating $\bm{\Omega} = \bS^{-1}$, often (but not always) using
low-rank structure and sparsity respectively. Example approaches that work with
$\bS$ include low-rank approximations~\parencite{cressie2008}, tapering-based methods
(a rare sparsity-based acceleration in ``$\bS$-space'')~\parencite{kaufman2008}, their
combination in the ``full-scale approximation'' (FSA)~\parencite{sang2012},
hierarchical matrix approximations that compress off-diagonal blocks~\parencite{ambikasaran2016,minden2017,litvinenko2019,geoga2020}, 
and reduced-operation methods that require only the action $\bm{v} \mapsto \bS
\bm{v}$~\parencite{anitescu2012,stein2013,ubaru2017,gardner2018}, which can then
additionally be accelerated with algorithms such as the fast multipole method
(FMM)~\parencite{greengard1987}.

Approximation methods that work directly with $\bm{\Omega} = \bS^{-1}$, on the
other hand, are more thematically uniform in that they effectively all make
Markovian-like approximations and produce sparse approximations to $\bm{\Omega}$
using arguments of conditional independence. Such approximations have many
variations, including Gauss-Markov random fields~\parencite{rue2005},
nearest-neighbor Gaussian processes~\parencite{datta2016,finley2019}, and Vecchia
approximations~\parencite{vecchia1988,stein2004,katzfuss2021}.  In this work, we will
use the last term to refer to the approximation.

Vecchia-type approximations are based on a simple idea about approximating
conditional distributions. In particular, observing that any joint distribution
can be expanded into a product of conditionals as
\begin{equation}\label{eq:factorization}
  f(y_1, \dots, y_n) = f(y_1) \prod_{j=2}^n f(y_j\ |\ y_1, \dots, y_{j-1}),
\end{equation}
Vecchia~\parencite{vecchia1988} proposed that for each conditional one could select a
subset of prior indices $\sigma(j) \subset [j-1] = \{1,2, \dots ,j-1\}$, typically
of size $|\sigma(j)| = \bO(1)$, and approximate the true density with
\begin{equation}\label{eq:approx}
  f(y_1, \dots, y_n) \approx f(y_1) \prod_{j=2}^n f(y_j\ |\ \{y_k:k\in\sigma(j)\}).
\end{equation}
This simple idea has several very favorable properties: it corresponds to a
valid multivariate density, it is embarrassingly parallel to evaluate, and if 
$|\sigma(j)| = \mathcal{O}(1)$ then the approximation is easily seen to only
require $\bO(n)$ work and $\bO(1)$ storage to evaluate. This method also has a
naturally implied sparse approximation to $\bm{\Omega} = \bS^{-1}$, and we refer
readers to~\parencite{katzfuss2021} for details and derivations.

Unlike ``$\bS$-space'' approximation methods that can be made adaptive, however,
Vecchia-type approximations are traditionally non-adaptive, and several design
choices can significantly impact their accuracy. The first design choice to make
is the order in which to enumerate measurements. In one dimension there is a
natural ordering to locations based on sorting, but in two or more dimensions
there is not a natural or canonical ordering. A second design choice and
challenge is that conditioning sets must be selected to specify the
approximation before optimizing, and typically stay fixed for the duration of
the optimization (with a notable exception being~\parencite{kang2023}). For processes
with complex dependence structure, it is not always clear how to design and
select conditioning sets. Considering that a typical size of $|\sigma(j)|$ is
around $10$ to keep runtime costs favorable (as runtime costs for each term will
grow with the familiar $\bO(|\sigma(j)|^3)$), poor choices for conditioning set
elements can be very costly in terms of efficiency.  With that said, an
increasingly large body of work shows Vecchia approximations emerging as a
leader in the broad field of approximation methods. For spatial data,
theoretically-motivated point ordering and conditioning set selection methods
have been proposed that demonstrate favorable
performance~\parencite{guinness2018,schafer2021}, and for slowly growing or even
fixed conditioning set sizes favorable asymptotic results have also been
given~\parencite{kang2023,szabo2024}.  Empirically, the approximation can out-compete
even much more expensive methods~\parencite{heaton2019}.

\begin{figure}!ht]
  \centering
  \includegraphics[width=0.25\textwidth]{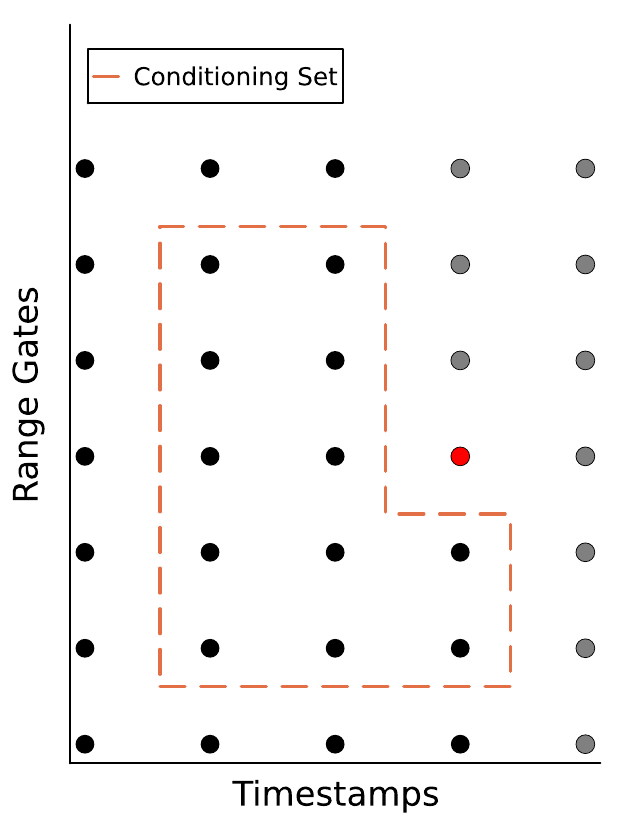}
  \caption{An example of selecting the conditioning set of the red point, 
    where the black points are previous ones, and the gray points have
    higher indices in the enumeration.}
  \label{fig:cond_set}
\end{figure}

In this work, however, where the data can effectively be viewed as a
vector-valued time series at hundreds of thousands of time points and tens of
altitudes, we opt for a simpler design approach than the state-of-the-art in
purely spatial data. In particular, we opt for a point ordering that is
``space-first'', and we select the conditioning set for measurement $y_j$ based
on prior enumerated spatio-temporal nearest neighbors in the intuitive way, as
exemplified in Figure~\ref{fig:cond_set}.

Yet another advantage of Vecchia approximation is that it is much easier than
many of its competitors to differentiate. Covariance function-based Gaussian
process models are notoriously difficult to fit, as it is commonly the case in
fixed-domain asymptotic regimes that only certain nonlinear functions of
parameters are consistently estimable and log-likelihood surfaces can exhibit
complex structure that is difficult for optimization routines, particularly
derivative-free methods, to navigate~\parencite{stein2013,geoga2023}. To address this
challenge, it is common to optimize using gradients, or ideally both gradients
and Hessians (or gradients and expected Fisher information matrices, which can
be very effective Hessian approximations~\parencite{geoga2020,guinness2021}). 
The gradient and Hessian of $\ell(\bm{\theta})$ are given by
\begin{equation}\label{eq:grad_hess}
  \begin{split}
    -2 \frac{\partial}{\partial \theta_j} \ell(\bth) =\ & \mathrm{tr} \fp{\bm{\Sigma}(\bm{\theta})^{-1} \bm{\Sigma}_j(\bm{\theta})}
    - \bm{y}^T \bm{\Sigma}(\bm{\theta})^{-1} \bm{\Sigma}_j(\bm{\theta}) \bm{\Sigma}(\bm{\theta})^{-1} \bm{y}, \\
    -2 \frac{\partial^2}{\partial \theta_j \partial \theta_k} \ell(\bth) =\ &
    \mathrm{tr} \fp{\bm{\Sigma}(\bm{\theta})^{-1} \bm{\Sigma}_j(\bm{\theta}) \bm{\Sigma}(\bm{\theta})^{-1} \bm{\Sigma}_k(\bm{\theta})} 
    - \mathrm{tr} \fp{\bm{\Sigma}(\bm{\theta})^{-1} \bm{\Sigma}_{jk}(\bm{\theta})} \\
    & + \bm{y}^T \fp{\frac{\partial}{\partial \theta_k} \fb{\bm{\Sigma}(\bm{\theta})^{-1} \bm{\Sigma}_j(\bm{\theta}) \bm{\Sigma}(\bm{\theta})^{-1}}} \bm{y},
  \end{split}
\end{equation}
where $\bm{\Sigma}_j(\bm{\theta}) = \frac{\partial}{\partial \theta_j}
\bm{\Sigma}(\bm{\theta})$ and $\bm{\Sigma}_{jk}(\bm{\theta}) =
\frac{\partial^2}{\partial \theta_j \partial \theta_k}
\bm{\Sigma}(\bm{\theta})$ are the first and second partial derivative matrices
of $\bS$ with respect to kernel parameters respectfully. In the case of Vecchia
approximations, these precise formulae may be directly applied term-by-term to
the log of~\eqref{eq:approx} to obtain the gradient and Hessian respectively.
This completely sidesteps the challenge many ``$\bS$-space'' methods face of
having to compute or approximate large matrix-matrix products involving the
(compressed) $n \times n$ matrix $\bS$.

As a final small wrinkle, we remark that effective closed-form derivatives of
$K$ with respect to all parameters are not available in closed form. Recent work
utilizing automatic differentiation (AD) has made estimating the smoothness 
parameter $\nu$ of the Mat\'ern model as easy as any other parameter~\parencite{griewank2008,geoga2023}, 
and this AD-based approach naturally can be extended to
many-parameter models such as this one, making it feasible to rapidly evaluate
derivatives and mixed second derivatives of the log-likelihood with respect to
even hundreds of distinct parameters. An important but subtle design choice of
this model is that parameters pertaining to the weight function $w(t, x)$ or the
ABL height function $\alpha(t)$ do not involve derivatives of the covariance
functions $K_1$, $K_2$, or $K_{12}$, which are by far the most expensive
functions in this model to programmatically differentiate, and makes it possible
to use {reverse-mode} AD (as opposed to the {forward-mode} AD used in
differentiating $\mathcal{K}_{\nu}$ in~\parencite{geoga2023}) to more efficiently
differentiate with respect to potentially hundreds of parameters in the basis
expansion of $\alpha(t)$. While this does not affect the asymptotic cost of the
differentiation, the prefactor improvement is significant.

We now discuss the specifics of the estimation problem in detail, noting that 
the optimal approach for retrieving ABL height varies with the desired temporal 
and spatial resolution. In the
case where one has a reasonably small number of parameters in $\alpha(t)$, say,
$m \lesssim 10$, one can simply use jointly optimize over both the ABL and
kernel parameters at once in a straightforward way using Vecchia
approximations and derivative methods discussed above. In the alternative case
where one is interested to obtain a more highly resolved estimate for the ABL
height, however, several modifications for scalability are necessary that we
will now outline in the next section.

\subsection{High-resolution ABL height retrieval}\label{sec:est_highres}

When selecting the family of functions $\{\psi_j\}$ to expand the ABL height
function $\alpha(t) = \sum_{j=1}^m c_j \psi_j(t)$ on, several standard options
come to mind: splines~\parencite{wahba1990} are a popular and well-studied tool for function
approximation, and so one could pick knot placements and obtain a commensurate
number of functions $\{\psi_j(t)\}_{j=1}^m$. One might also consider Chebyshev
or some other orthogonal polynomial family~\parencite{trefethen2019}. Both of those choices, and
surely many others, are perfectly good ideas, and depending on the application
may be the best choice.  In the case where $m$ is large, however, a particularly
elegant option is to use the domain partitioning and machinery of wavelets~\parencite{mallat1999}, 
which provide compactly supported and orthogonal functions. By
exploiting the dyadic support structure and orthogonality of the shifted and
scaled wavelet functions, one can obtain high-quality estimates for
$\{c_j\}_{j=1}^m$ for very large $m$ by solving successive optimization problems
of constant dimension.

\begin{figure}[!ht]
  \centering
  \includegraphics[width = 0.8\textwidth]{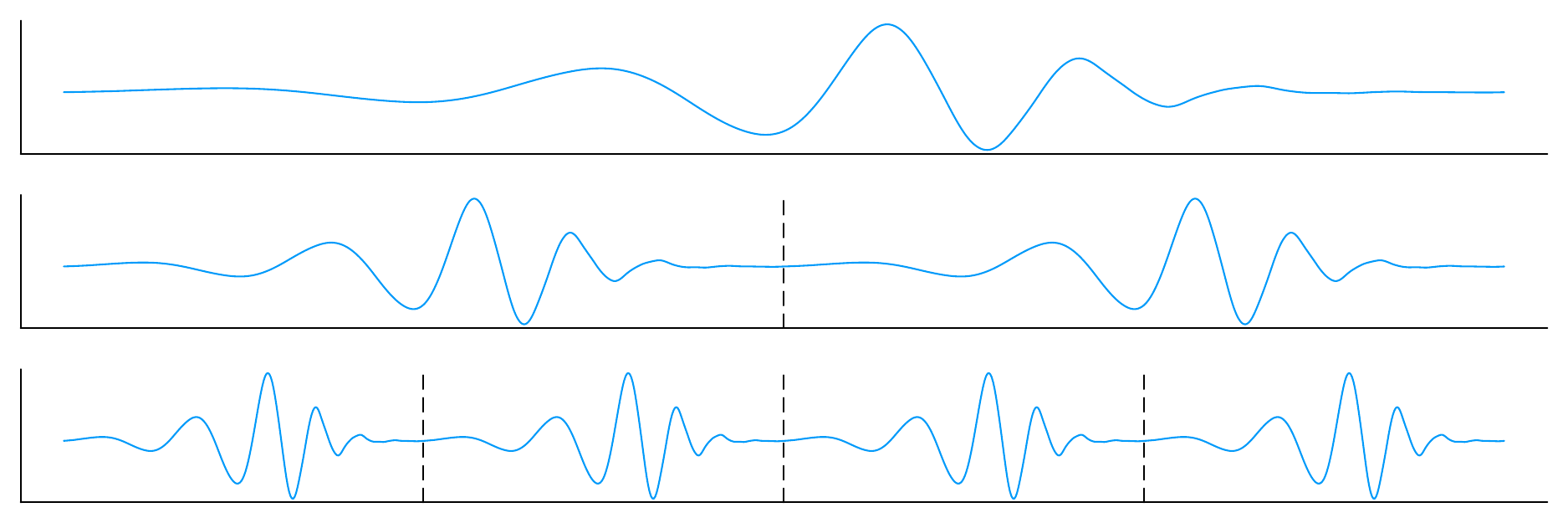}
  \caption{Daubechies wavelets of order 8 at three different resolution levels.}
  \label{fig:wavelets}
\end{figure}

A core concept in wavelet theory is a {multiresolution approximation}
(MRA), which is a sequence $\{V_i\}_{i \in \mathbb{Z}}$ of closed subspaces of
$L^2(\mathbb{R})$ indexed by frequency resolution satisfying, among other
things, the fact that for $f(t) \in V_i$ one has that (i) translations 
$f(t - 2^i k)$ are also in $V_i$ and (ii) scalings $f(t/2) \in V_{i + 1}$
(see~\parencite{mallat1999} for a full definition and expository introduction).
Oftentimes, MRAs are generated by the scaling and translation of a 
compactly supported {scaling function} or {mother wavelet}, $\psi(x)$, so 
that $f(t) \in V_i$ is of the form $\psi(2^{-i} t - k)$ for some $i$ and 
$k$. Figure~\ref{fig:wavelets} shows an example of the Daubechies wavelet 
of order $8$~\parencite{daubechies1992} at three increasing levels, with vertical lines indicating the endpoints of support for each individual element in $V_i$.

As it pertains to this work, we introduce the following notation. Assuming 
that we are working with a time segment supported on an interval $[0, 2^I]$, 
we pick a mother wavelet $\psi(t)$ supported on $[0, 1]$ at level $i = 0$, 
so that the support of an element in $V_I$ is precisely $[0, 2^I]$. Let that
element be denoted $\psi_{I, 0}$. In $V_{I - 1}$, then, there will be two
elements supported on intervals of length $2^{I - 1}$ that are within 
$[0, 2^I]$, which we denote $\{\psi_{I - 1, 0}, \psi_{I - 1, 1}\}$.
Iteratively repeating this process then gives the collection of functions 
$\{ \{\psi_{i, k}\}_{k = 0}^{2^{I - i} - 1} \}_{i = 0}^I$.
Using a lexicographic enumeration, we simplify the indexing to give
$\{\psi_j\}_{j=1}^m$, where $m = 2^{I + 1} - 1$. In this notation, we note 
that the functions plotted in Figure~\ref{fig:wavelets} can be interpreted 
as $\psi_1$ in the first row, $\psi_2$ and $\psi_3$ in the second, and 
$\{\psi_j\}_{j = 4}^7$ in the third row.

\begin{figure}[!ht]
  \centering
  \includegraphics[width = 0.8\textwidth]{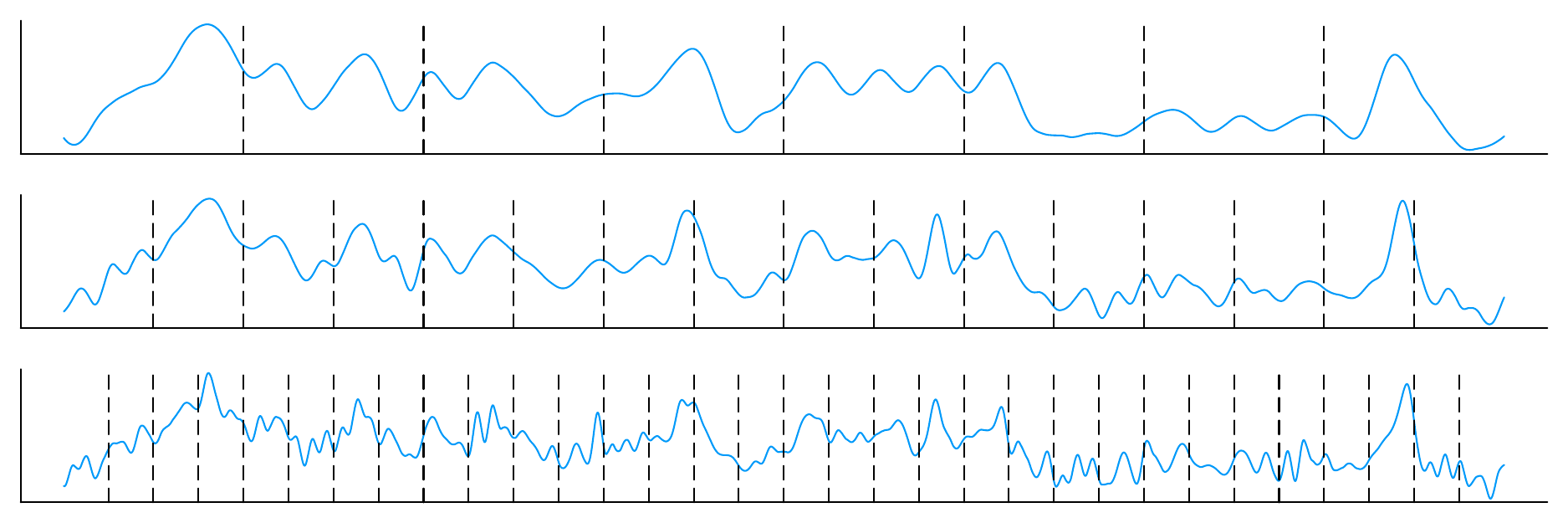}
  \caption{The evolution of a signal with wavelets from low to high 
  frequencies with vertical lines indicating the support of functions $\psi_j$
  as the level is decreased.}
  \label{fig:signal}
\end{figure}

The value in this choice of expanding 
$\alpha(t) = \sum_{j=1}^m c_j \psi_j(t)$ is that, given fixed covariance 
parameters (which can be estimated in advance in a process that will be 
detailed in the next section), one can successively refine the estimated 
function $\alpha(t)$ by decreasing $i$, and within each level $i$ each 
parameter can be estimated independently as their supports are disjoint. 
This refinement like process yields sequentially more highly-resolved 
estimates of the true ABL height, exemplified with synthetic data in 
Figure~\ref{fig:signal}. This procedure is summarized in 
Algorithm~\ref{alg:est_wavelet}, with small details like penalizing for
$\{c_j\}_{j = 1}^m$ and the specifics of initialization omitted for clarity.

\begin{algorithm}
  \DontPrintSemicolon
  \SetArgSty{textnormal}
  
  \KwData{initial coarse estimate for $\alpha(t)$}
  \KwResult{
  $\alpha(t)$ with optimized wavelet coefficients 
  $\{ \{\hat{c}_{i, k}\}_{k = 0}^{2^{I - i} - 1} \}_{i = 0}^I$
  }
  
  \tcp{Loop over each wavelet level, starting at the coarsest}
  \For{$i \gets I$ \KwTo $0$}{  
    \tcp{Independently fit coefficients for each disjointly supported $\psi_{i, k}$}
    \For{$k \gets 0$ \KwTo $2^{I - i} - 1$}{
      \tcp{Estimate the coefficient of $\psi_{i, k}$ by minimizing the negative log-likelihood}
      $\hat{c}_{i, k} \gets \arg\min -\ell(\alpha(t) + c_{i, k} \psi_{i, k}(t))$\;
      \tcp{Update the coefficient of $\psi_{i, k}$ in $\alpha(t)$}
      $\alpha(t) \gets \alpha(t) + \hat{c}_{i, k} \psi_{i, k}(t)$
    }
  }
  \caption{Optimization of wavelet coefficients.}
  \label{alg:est_wavelet}
\end{algorithm}

Algorithm~\ref{alg:est_wavelet} updates a single coefficient at 
each iteration of the inner loop. In this setting, a 
derivative-free optimization method may be employed to estimate 
$c_{i, k}$, thereby reducing computational cost by avoiding the 
evaluation of derivatives and Hessians of the log-likelihood. 
Alternatively, multiple coefficients may be optimized 
simultaneously to reduce the total number of inner-loop 
iterations; however, this approach typically requires 
derivatives and Hessians to achieve adequate computational 
efficiency. In practice, one can compare the performance of 
these strategies and choose the most effective design.

\section{Demonstration of ABL height estimation}\label{sec:results}
In this section, we demonstrate the application of the proposed GP framework to 
the retrieval of ABL height from DL vertical velocity profiles. We present two 
case studies. First, we illustrate an hour-long, high temporal resolution retrieval 
to highlight the model's ability to resolve fine-scale ABL evolution. Second, we 
examine full-day retrievals under varying meteorological conditions, including
periods with cloud contamination where reliable lidar returns are unavailable. 
We further provide summaries of the covariance function parameters, demonstrating 
how the model leverages different local properties of the process above and below
the ABL height to accurately distinguish between turbulent and non-turbulent 
regimes across fine timescales. All data analyzed in this section are provided 
by the Department of Energy's CROCUS Urban Integrated Field Laboratory project~\parencite{muradyan2025crocusmicronet, collis2025crocusurbancanyons}.

\subsection{High-resolution ABL recovery}\label{sec:high_res}

\begin{figure}[!ht]
  \centering
  \includegraphics[width = 0.8\textwidth]{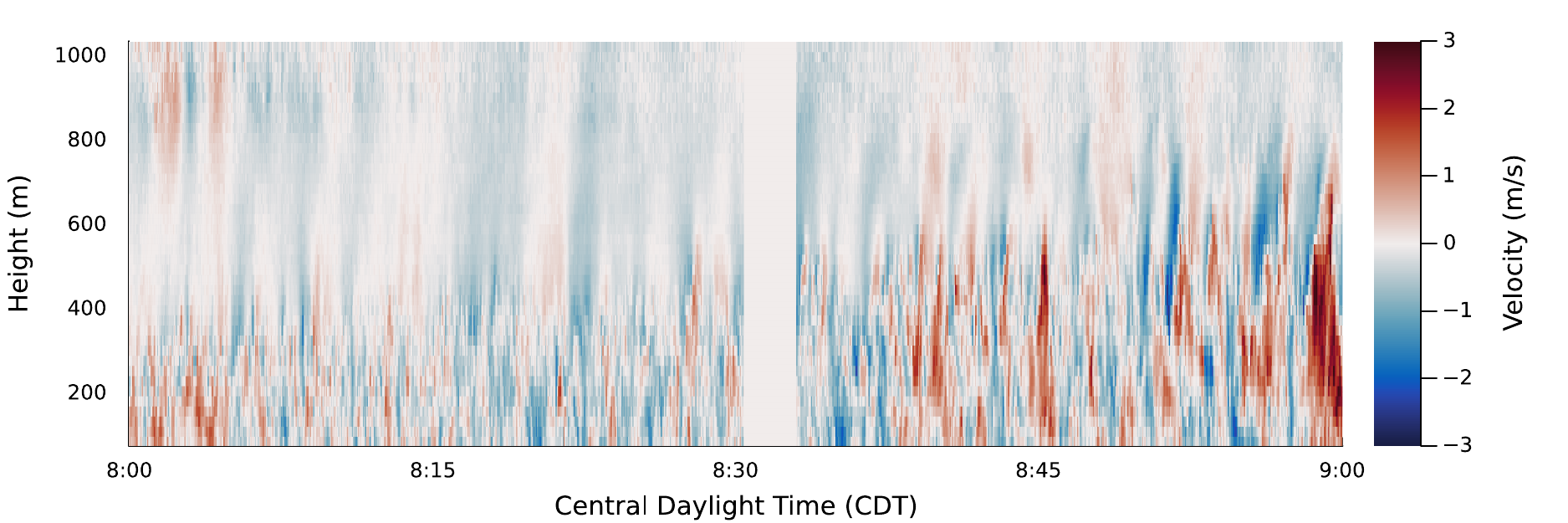}
  \caption{Heatmap of the vertical wind velocity from 8 AM to 9 AM (CDT) on July 20, 2023.}
  \label{fig:one_hour_origin}
\end{figure}

As an illustrative case, we analyze DL measurements collected from 8 AM to 9 AM 
(CDT) on July 20, 2023, consisting of vertical velocity observations at 40 range 
gates with 24-m spacing. Figure~\ref{fig:one_hour_origin} shows how the vertical 
velocity varies with time and height, and the gap at approximately $8\text{:}30$ 
AM corresponds to vertical stare measurement interruption as the instrument executes 
a horizontal scanning sequence to produce a profile of horizontal wind velocities. 
This morning period is of particular meteorological interest, as surface heating 
due to solar radiation initiates convective mixing, driving rapid growth of the 
developing convective boundary layer. Retrieving the ABL height at high temporal 
resolution involves three key steps: (1) estimating covariance parameters, (2) 
obtaining an initializer for the ABL height, and (3) successively refining it as 
described in Section~\ref{sec:est_highres}.

\begin{figure}[!ht]
  \centering
  \includegraphics[width = 0.8\textwidth]{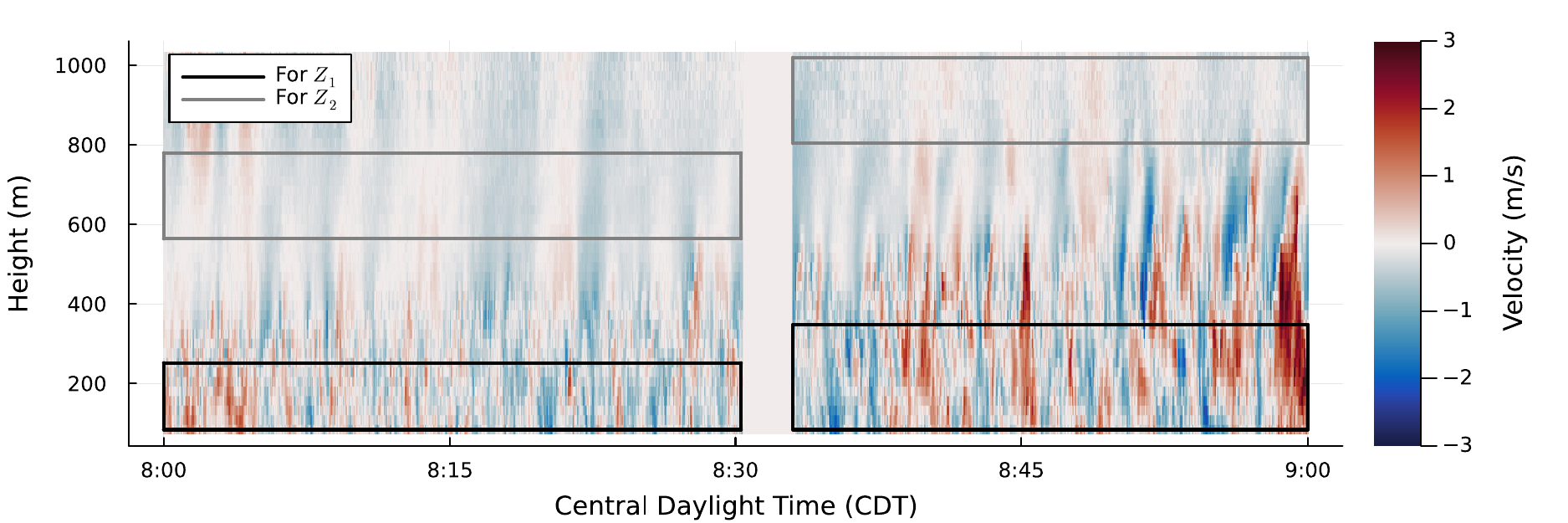}
  \caption{Areas for calculating the MLE of the covariance parameters of $Z_1$ and $Z_2$ on July 20, 2023.}
  \label{fig:one_hour_squares}
\end{figure}

For the first step, covariance parameters of model~\eqref{eq:model} are estimated 
using data from subsets of the domain near the lowest and highest altitudes where 
the instrument SNR exceeds the quality-control threshold~\parencite{newsom2022doppler}. 
We then fit the bivariate Mat\'ern model of~\eqref{eq:model} to that data, assuming 
that these regions are entirely within or above the ABL, such that the weight 
$w(t,x)$ is either zero or one to effectively machine
precision. Figure~\ref{fig:one_hour_squares} shows the areas for calculating
the MLE of the covariance parameters of $Z_1$ and $Z_2$. The two areas enclosed
in black rectangles for the parameters of $Z_1$, i.e. $\bth_1 = (\sigma_1,
\rho^t_1, \rho^x_1, \nu_1)$, where the nugget variance $\tau_1$ is excluded
since anecdotally with many data segments the optimizer consistently selects it
to be zero. The gray boxes indicate data used to determine $\bth_2 = (\sigma_2,
\rho^t_2, \rho^x_2, \nu_2, \tau_2)$, the parameters corresponding to the
covariance function for $Z_2$. Estimated parameter values are summarized in Table~\ref{tab:cov_mle}.

Several aspects of these results are worth noting. First and most interestingly, 
the smoothness parameter $\nu$ is estimated to be lower above the ABL than in it, 
suggesting that the process aloft is modeled as rougher in space and time, contrary 
to the conventional expectation that turbulent motions within the ABL exhibit 
greater small-scale variability~\parencite{lenschow1974model, stull2012}. This behavior 
may reflect the limited SNR aloft, which can lead the MLE to attribute additional 
variability to the process rather than the nugget term. Interestingly, the estimated 
nugget variance above the ABL of $\approx 0.0025$ is smaller than expected, 
indicating that the optimizer prefers to describe $Z_2$ as a rough but low-noise 
process rather than a smooth and noisy one. The remaining parameters follow expected 
trends: marginal variance and temporal correlation length scale are larger and 
shorter respectively, for $Z_1$, consistent with higher variability and shorter 
decorrelation times characteristic of the turbulent ABL. The spatial (vertical) 
correlation above the ABL is more pronounced, reflecting the smoother, more coherent 
flow typical of the free atmosphere. 

\begin{figure}[!ht]
    \centering
    \begin{tabular}{c}
      \includegraphics[width=0.8\linewidth]{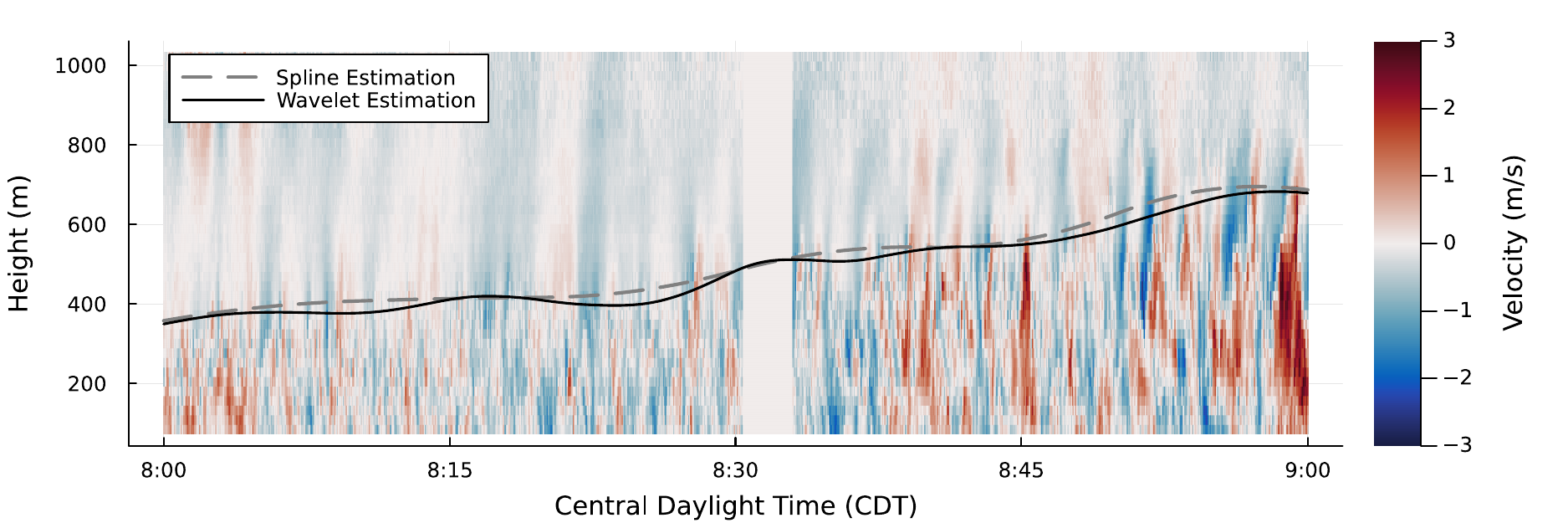} \\
      \includegraphics[width=0.8\linewidth]{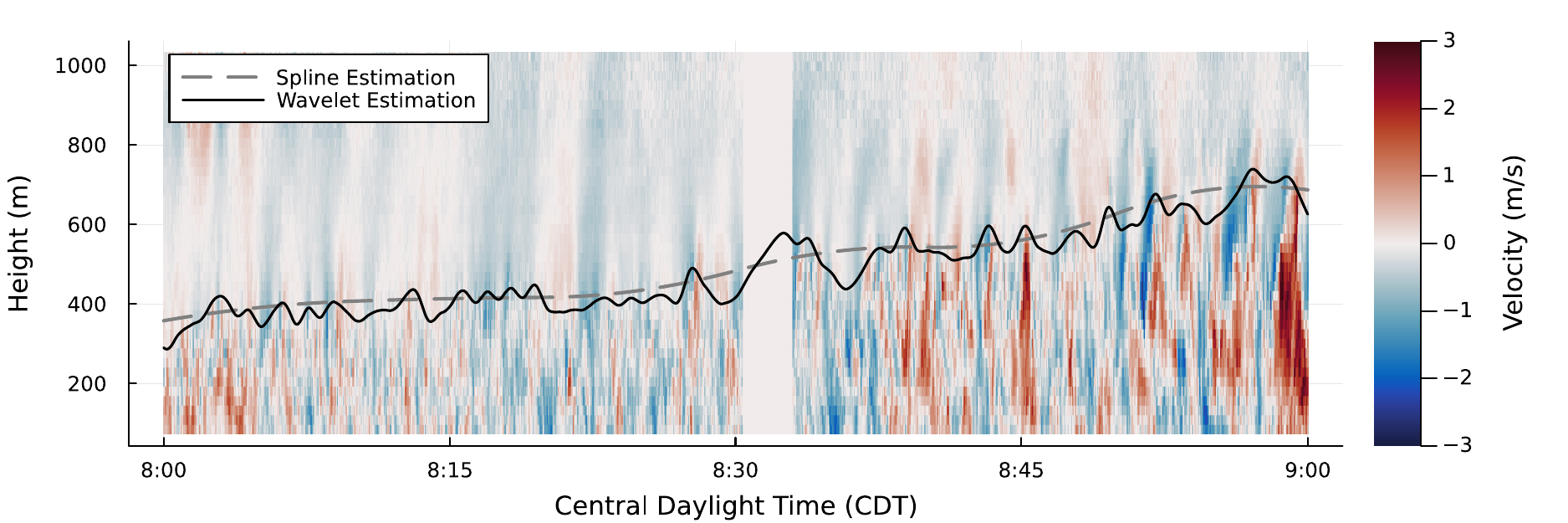} \\
      \includegraphics[width=0.8\linewidth]{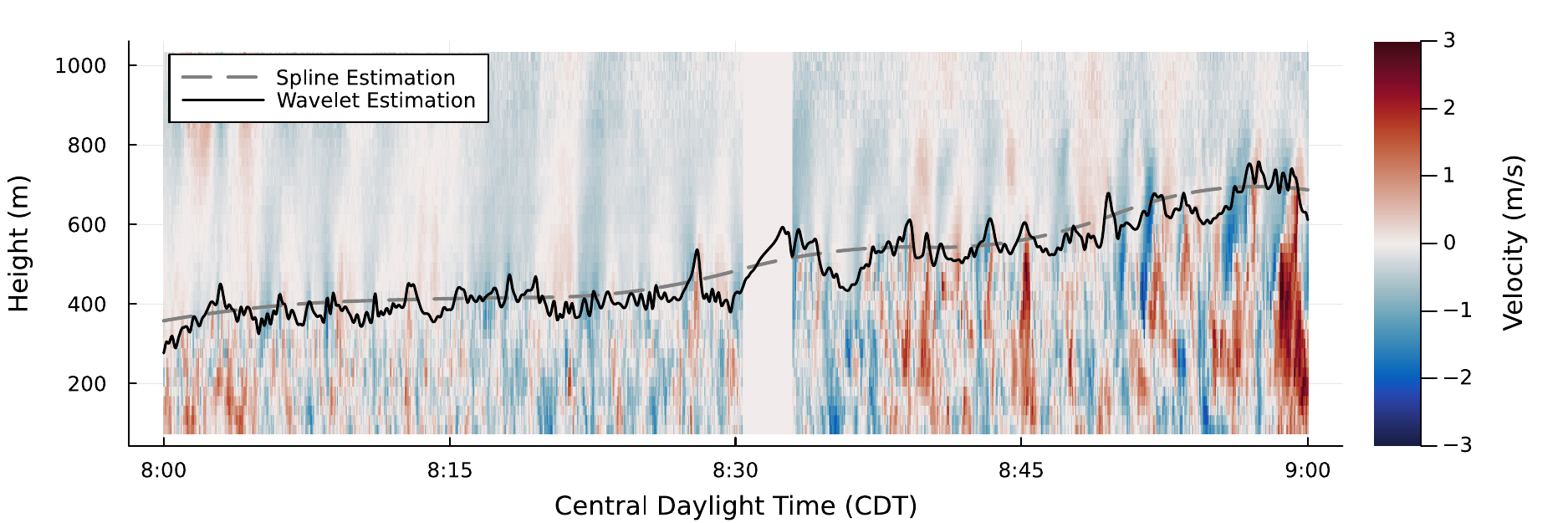}
    \end{tabular}
    \caption{The ABL height estimation using wavelets on July 20, 2023 using
    increasingly fine wavelet resolutions for the expansion of $\alpha(t)$. In
    all panels, the spline initializer is shown in grey.}
    \label{fig:one_hour_wavelet}
\end{figure}

With $\widehat{\bth}_1$ and $\widehat{\bth}_2$, we now fit a coarse spline basis
with eight knots, again using the log-likelihood, to initialize the ABL height.
Successive refinements of this spline-based initializer are then computed. The
results of multiple levels of refinement are summarized in Figure~\ref{fig:one_hour_wavelet}. 
The dashed gray line in each subplot shows the spline-based initializer. The 
solid black line is the estimated ABL height using Daubechies wavelets, which 
in each row of Figure~\ref{fig:one_hour_wavelet} adds two finer levels of resolution, 
totaling to $m = 384$ wavelets in the final figure. With each refinement, this spline
estimator can be observed to accurately capture new fine scale features: moving
from the first panel to the second, for example, we note that the dip in ABL
height at approximately $8\text{:}35$ AM is captured, and that moving from the
second to the third panel several sharp gusts---such as at approximately
$8\text{:}25$ and $8\text{:}50$ AM---are now captured. 

\subsection{Full-day ABL profiles}\label{sec:full_day}

In this section, we reduce the temporal resolution of the ABL height retrieval 
function $\alpha(t)$ to produce full-day profiles that capture the complete 
diurnal cycle. The key additional challenge in this setting is that the process 
is not stationary in time, as the underlying turbulence, stability, and aerosol 
structure of the ABL vary systematically with progression from the convective 
boundary layer to the stable nocturnal layer~\parencite{stull2012, garratt1994, mahrt2014stably}. 
As a result, the covariance parameters $\bm{\theta}$ must be re-estimated 
adaptively as the time of day changes. This estimation procedure largely follows 
the high-resolution case described in the previous section, with several important 
modifications. First, a temporal smoothness constraint implemented as a small 
continuity penalty at the endpoints of consecutive hourly segments is applied 
to ensure piecewise estimates for $\alpha(t)$ remain continuous across the full 
24-hour period. Second, the covariance parameters are reinitialized each
hour using the in- or above-ABL process parameters inferred from the preceding 
hour. This allows the model to adapt to gradual diurnal evolution in stability 
and turbulence intensity.

\begin{figure}[!ht]
  \centering
  \begin{tabular}{c}
    \includegraphics[width = 0.8\textwidth]{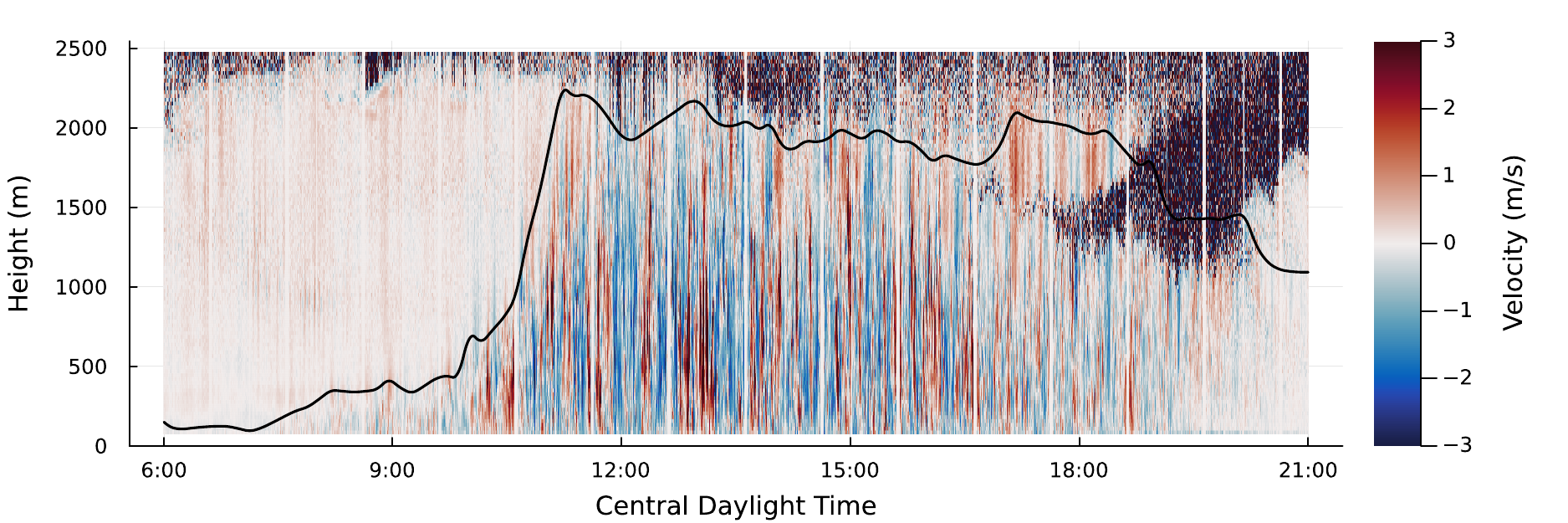} \\
    \includegraphics[width = 0.8\textwidth]{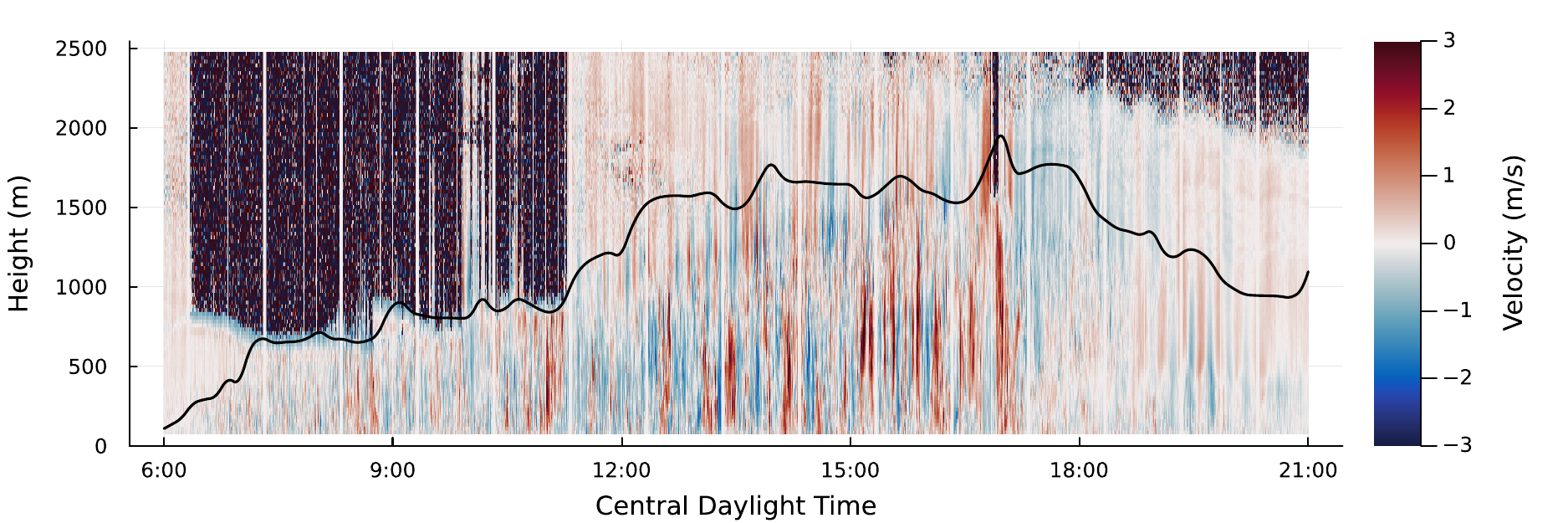} \\
    \includegraphics[width = 0.8\textwidth]{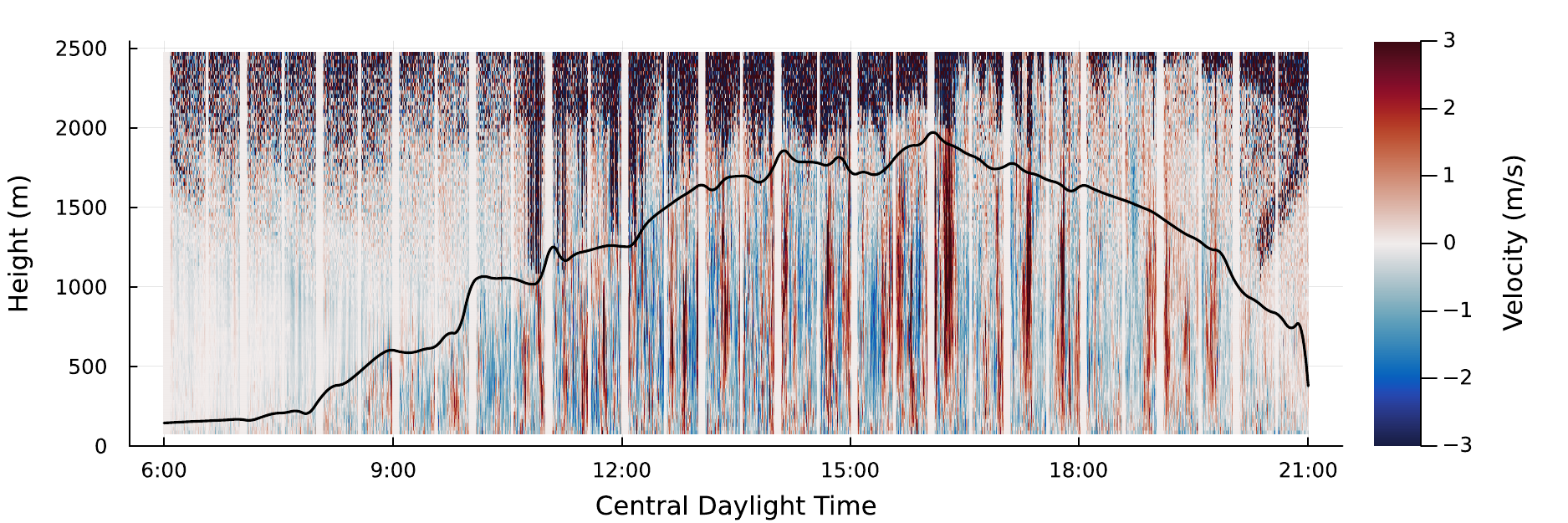}
  \end{tabular}
  \caption{The ABL height estimation using splines on June 23, 2023 (top row),
  June 27, 2023 (middle row), and July 20 (bottom row).}
  \label{fig:full_day_profiles}
\end{figure}

We analyze three days with distinct meteorological conditions and cloud regimes. 
For each day, data are available from 6 AM to 9 PM local time, spanning the full 
period from morning transition to evening stabilization. For each hour, we use a 
simple linear function as the initial guess and six quadratic splines to estimate 
the ABL height. After concatenating all hourly fits, we use 60 quadratic splines 
to smooth the contact points of the ABL height of different hours. Figure~\ref{fig:full_day_profiles} 
summarizes the estimated ABL height for these three days, selected to exhibit 
different patterns of data availability and atmospheric variability.

As Figure~\ref{fig:full_day_profiles} demonstrates, the diurnal evolution of the 
ABL height is captured realistically across diverse meteorological conditions, 
including periods of missing or low-SNR data and transitions between convective, 
residual and stable regimes. The three examples exhibit distinct temporal correlation 
structures: June $27$ (middle panel) displays the longest dependence scales and 
smoothest ABL height growth, consistent with weaker convective forcing and 
intermittent cloud shading (clouds between 6:15 and 11:40 AM), whereas July $20$ 
(bottom panel) shows strong short-term variability and enhanced turbulence 
intermittency, indicative of vigorous convective mixing likely under clear-sky, 
high insolation conditions. In contrast, June $23$ (top panel) represents a more 
transitional case, characterized by a well-defined growth phase followed by gradual 
afternoon stabilization. On July $20$, the ABL remains deep and coherent throughout 
most of the day, reflecting persistent surface heating and sustained convective 
overturning, with the ABL height oscillations after local noon likely associated 
with entrainment-zone variability and thermal plume organization rather than 
retrieval noise. These examples collectively highlight that the proposed GP 
retrieval framework performs robustly across a range of atmospheric states, 
maintaining physically consistent ABL height estimates even under strong turbulence, 
partial cloud cover, or transient data gaps. Moreover, the differences among the 
three days underscore that different covariance features (e.g., marginal variance, 
temporal correlation length, or nugget variance) can dominate as discriminants 
between in- and above-ABL regimes, depending on the prevailing meteorological 
forcing and turbulence intensity.

Finally, it is worth noting that because the proposed model is globally 
nonstationary in time but locally stationary within short intervals, it naturally 
adapts to the diurnal transition between boundary layer regimes. The mixture model 
for $Z(t, x)$, representing distinct dynamic behaviors in and above the ABL, allows 
the same framework to continuously transition from the convective daytime boundary 
layer to the stable nocturnal boundary layer (NBL) without any explicit change in 
model structure or thresholds. This capability is particularly valuable, as 
retrieving the NBL height remains a persistent challenge for traditional 
variance-threshold or gradient-based methods, owing to the weaker turbulence, 
smaller vertical gradients, and reduced contrast between sub- and super-layer 
properties at night~\parencite{van2012cessation, mahrt2014stably}. By capturing these 
gradual transitions through time-varying covariance parameters, the GP approach 
provides a physically consistent and automated means of identifying both convective 
and stable boundary-layer regimes within a single probabilistic framework.

\section{Discussion}\label{sec:dis}
High temporal resolution retrievals of the ABL height offer significant advances 
for both scientific understanding and operational applications. The ABL height 
as a function of time characterizes the dynamic boundary between the surface and 
the free atmosphere. Traditional hourly estimates can miss the rapid fluctuations 
of this boundary, particularly during morning growth, evening collapse, or episodic 
events such as lake breezes, convection bursts, or frontal passages, when changes 
can occur on timescales of minutes. Resolving these short-term fluctuations enables 
better quantification of entrainment processes at the top of the boundary layer, 
improved closure of energy and mass budgets, and more accurate representation of 
turbulence intermittency. For researchers, such fine-scale ABL height observations 
provide essential validation targets for high-resolution models, large-eddy 
simulations, and land-atmosphere coupling studies. Operational forecasters, 
regulatory and environmental agencies such as the National Weather Service and 
the U.S. Environmental Protection Agency, as well as energy and transportation 
sectors, can all benefit from rapid updates in boundary-layer depth. Such 
information improves real-time assessments of surface pollutant concentrations, 
plume dispersion, fog and low-cloud formation, and the timing of convection 
initiation. Beyond the research community, urban planners, public-health officials, 
and emergency-response organizations can also leverage these data for exposure 
risk evaluation, smoke and hazardous-release dispersion modeling, near-surface 
temperature and humidity forecasting, and applications where the difference between 
a one-minute and one-hour temporal resolution can critically influence operational 
decisions and public-safety outcomes.

In this study, we introduced a method for using purpose-built Gaussian process
models to identify the height of the atmospheric boundary layer using both
marginal- and cross-covariance structure of the process within and above it.
Unlike prior work in this area~\parencite{geoga2021}, this method allows a fully varying ABL
height $\alpha(t)$ and can be used to obtain extremely high-resolution estimates
for the time-varying process. With that said, however, many important
statistical questions remain: a more careful analysis of the appropriate level
of shrinkage used to avoid introducing spurious fluctuations in $\alpha(t)$
would be valuable, and a more flexible model for the marginal- and
cross-dependence structure of the $Z_1$ and $Z_2$ processes may result in better
estimates. Further, the model requires having {some} data both within
and above the ABL, and thus it is not viable in situations such as calm, stable nights when the nocturnal ABL becomes too shallow to be sampled. These questions and many application-based enhancements are
exciting questions for future work.

\printbibliography

\newpage

\begin{table}[!ht]
  \centering
  \footnotesize
  \renewcommand{\arraystretch}{1.5}
  \begin{tabular}{|l|l|l|l|l|l|}
    \hline
    \multicolumn{1}{|c|}{GP} & \multicolumn{1}{c|}{$\widehat{\sigma}$} & \multicolumn{1}{c|}{$\widehat{\rho^t}$} & \multicolumn{1}{c|}{$\widehat{\rho^x}$} & \multicolumn{1}{c|}{$\widehat{\nu}$} & \multicolumn{1}{c|}{$\widehat{\tau}$} \\ 
    \hline 
    \multicolumn{1}{|c|}{$Z_1$} & 6.64e-1 (9.59e-3) & 2.62 (4.69e-2) & 9.12e1 (1.38) & 1.52 (1.44e-2) & \\ 
    \multicolumn{1}{|c|}{$Z_2$} & 4.30e-1 (1.08e-1) & 1.45e2 (5.79e1) & 1.24e3 (5.02e2) & 7.91e-1 (3.21e-2) & 5.50e-2 (2.18e-4) \\ 
    \hline
  \end{tabular}
  \caption{The MLE of parameters of $Z_1$ and $Z_2$ based on Figure~\ref{fig:one_hour_squares} with standard deviations provided in parentheses.}
  \label{tab:cov_mle}
\end{table}

\end{document}